\documentclass[%
	final,
	notitlepage,
	narroweqnarray,
	inline,
	twoside,
	]{ieee}

\usepackage{array,graphicx,float,epsfig,wrapfig,makeidx,psfrag,booktabs,tabularx}

\begin{document}

\title[Nanonewton force Generation and Detection]{%
       Nanonewton force generation and detection based on a sensitive torsion pendulum}

\author[]{Sheng-Jui Chen\authorinfo{The authors are with the Center for Measurement Standards, Industrial Technology Research Institute, 
       Hsinchu, Taiwan 300, R.O.C.  
       Corresponding e-mail address: SJ.Chen@itri.org.tw}%
\and{}and Sheau-Shi Pan}




\maketitle               

\begin{abstract} 

In this paper, 
we introduce the experiment based on a sensitive torsion pendulum for measuring and calibrating small forces at nanonewton scale.  
The force standard for calibration is the universal gravitation between four masses separated by known distances.  
It is realized by two test masses suspended as the part of torsion pendulum and two source masses on a rotation table.  
Two force generation mechanisms, optical force from radiation pressure and electrostatic force by capacitive actuation unit, 
are designed and will be calibrated by the gravitation force.  
We present our recent results of radiation pressure measurements, 
and describe the design of capacitive displacement sensing/actuating unit.  
 
\end{abstract}

\begin{keywords}
Torsion pendulum, radiation pressure, active damping control, capacitive accelerometer, nanonewton, gravitational constant.  
\end{keywords}

\section{Introduction}

\PARstart Generation and measurement of force in micro/nano newton level has many applications in various fields of technology.  
Typical applications include the atomic force microscopy and instrumented indentation for nanoscale material characterization.  
The NIST microforce project \cite{Newell} has demonstrated the SI force realization and measurement of force below 5 $\mu$N.  
The force is electrostatic force and linked to electrical unit standards, the Josephson and quantized Hall effects.  
Another group from PTB has just published their development concept of facility for measuring forces below 10 $\mu$N with a resolution of pico-newton \cite{Nesterov}.  
In the center for measurement standards (CMS), we have built an torsion pendulum facility for generating and detecting forces at nano newton scale.  
The apparatus is capable of producing force by three methods, the radiation pressure and electrostatic effects, and the gravitation.  
The last one is realized by four cylindrical masses.  
Four cylinders are arranged in a way as in experiments for measuring the gravitational constant G, 
which allows precisely known gravitation force to be produced.  
This horizontal gravitation force is taken as the force standard and used to balance forces produced from other two methods.  
In the following sections, we describe the experimental setup in detail, 
and present recent results of radiation pressure damping experiment \cite{Chen}.

\section{Experimental setup}
\begin{figure}
\begin{center}
\includegraphics[angle=0,scale=0.7]{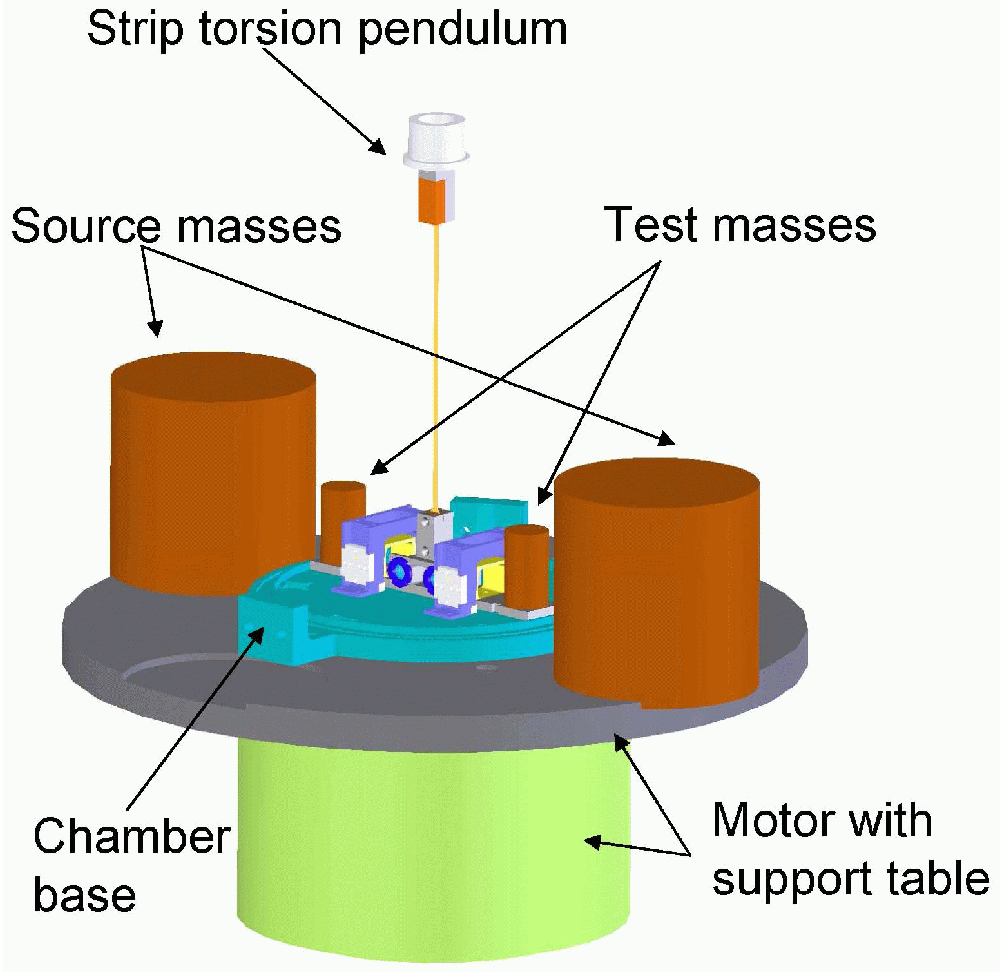}
\caption{Schematic view of the apparatus.}
\label{apparatus}
\end{center}
\end{figure}
The key component of the apparatus is a very thin, wide Cu-Be strip similar to that used in the experiment for measuring Newtonian constant of gravity at 
the Bureau International des Poids et Mesures (BIPM) \cite{Quinn_1}.  
It is 200 mm in length, 1.2 mm in width and 27 $\mu$m in thickness.  
The advantage of the strip, when heavily loaded, is that its restoring torque is mainly gravitational.  
Unwanted effects such as low frequency drift, 
anelastisity coming from the intrinsic properties of material can be neglected compared with gravitational signal \cite{Quinn_2}.  
The schematic view of the apparatus is shown in figure \ref{apparatus}.  
The torsion pendulum consists of the Cu-Be strip, an aluminum alloy frame and two cylindrical test masses.  
The aluminum alloy frame is suspended by the Cu-Be strip through a brass clamper.  
The torsion pendulum, under a total load of about 0.86 kg, has a spring constant of $5.25\times10^{-6}\;\mbox{N}\cdot\mbox{m}/\mbox{rad}$, 
and the resonant frequency for it is 5.3 mHz \cite{Wu}.  
In order to isolate noises from the environment, 
the pendulum is enclosed by a vacuum chamber with its interior pressure kept at several mTorr by a mechanical vacuum pump.  
Outside the vacuum chamber, another two copper cylindrical source masses are placed on a rotation table for producing gravitational torque.  
They each have a mass of about 18.4 kg.  The maximum gravitational torque is evaluated to be $2\times10^{-9}$ Nm.  
For armlength of 70 mm, the dynamic range for force calibration is below 28 nano newton.  

\begin{figure}
\begin{center}
\psfrag{diff}[][][1.5][0]{$\frac{d}{dt}$}
\includegraphics[angle=0,scale=0.5]{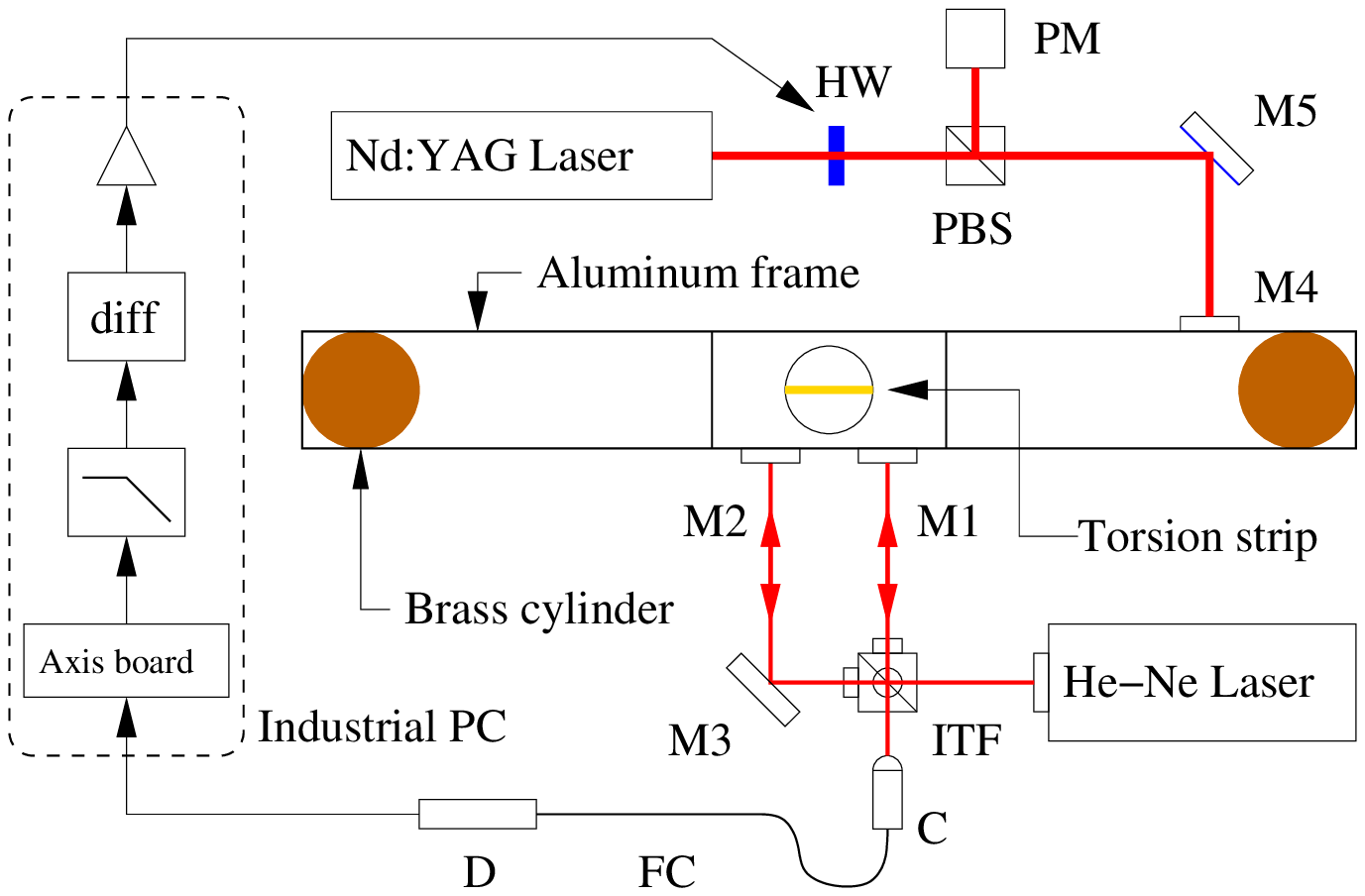}
\caption{Schematic view of experimental setup.  ITF: interferometer; M1 to M5:  reflection mirrors; C:  collimator; FC:  Fiber cable; D: photo-detector; 
PBS: polarizing beam splitter; PM: power meter.}
\label{ex_setup}
\end{center}
\end{figure}
The torsion pendulum has a quality factor $Q$ of about 272 under current vacuum condition.  
Figure \ref{ex_setup} shows the details of the torsion pendulum and the setup for radiation damping system.  
The deflection is detected by measuring the differential displacement between M1, M2 with a heterodyne interferometer ITF.  
The interferometer when incorporated with its processing electronics has a displacement resolution of 10 nm, 
which corresponds to an angular resolution of 0.39 $\mu$rad.   

\section[Recent results on radiation pressure force]{Radiation pressure force}

The phenomenon that photon exerts force on any surface hit by it is know as radiation pressure effect.  
This is the consequence of that photon possesses momentum of 
\begin{equation}
p=h/\lambda\;,
\end{equation} 
where $h$ is the Planck constant and $\lambda$ is the wavelength of light. 
For a surface with reflectivity of 1, the radiation force exerted upon it can be written as 
\begin{equation} 
f_{r}=\frac{2P}{c}\;\label{F_r}, 
\end{equation}
where $P$ and $c$ are the power and the speed of light respectively.  
This tiny force is becoming more visible and useful with advance in science and technology.  
As in laser interferometric gravitational wave detectors, 
it is proposed to use radiation pressure to reduce its thermal noise \cite{Cohadon}, 
or to actuate suspended test masses for interferometer control \cite{Feat}.  
Another recent interest is to cool down a mechanical resonator toward its quantum ground state by radiation pressure \cite{Metzger,Kleckner}.  

The radiation source is a 1064 nm Nd:YAG laser of nearly 1.5 W.  
The intensity of the Nd:YAG laser is modulated by an active variable attenuator.  
The attenuator is assembled by the power meter PM, the polarizing beamsplitter PBS and 
the half-wave plate HW mounted on a servo controlled rotary stage.  
The desired output power is achieved by adjusting the angle of the HW according to the output of the PM.  
This is done by a simple PID controller implemented on a industrial PC.  The bandwidth for this control is greater than 1 Hz.  
The mirror M4 attached on the pendulum provides a reflecting 
surface for momentum transfer between photons and the 
pendulum.  Its reflectivity is greater than 0.99.

\subsection{Damping control and force measurement}
Because the radiation pressure can only push the pendulum, 
to apply an effective pull force the incident power must be biased at a DC value.  
We set this bias DC value to be 0.68 W.  
The intensity of Nd:YAG laser is varied according to the measured deflection angle by a digital servo filter.  
The servo filter limits the bandwidth of feedback signal by two low pass filters whose cut-off frequencies are 0.01 Hz and 0.03 Hz respectively.  
The differential of the low-pass filtered signal is then fed back to the active variable attenuator with proper gain.  
In this way, the radiation force is mainly proportional to the pendulum's angular velocity, and creates damping effect.  

\begin{figure}
\begin{center}
\includegraphics[angle=0,scale=0.63]{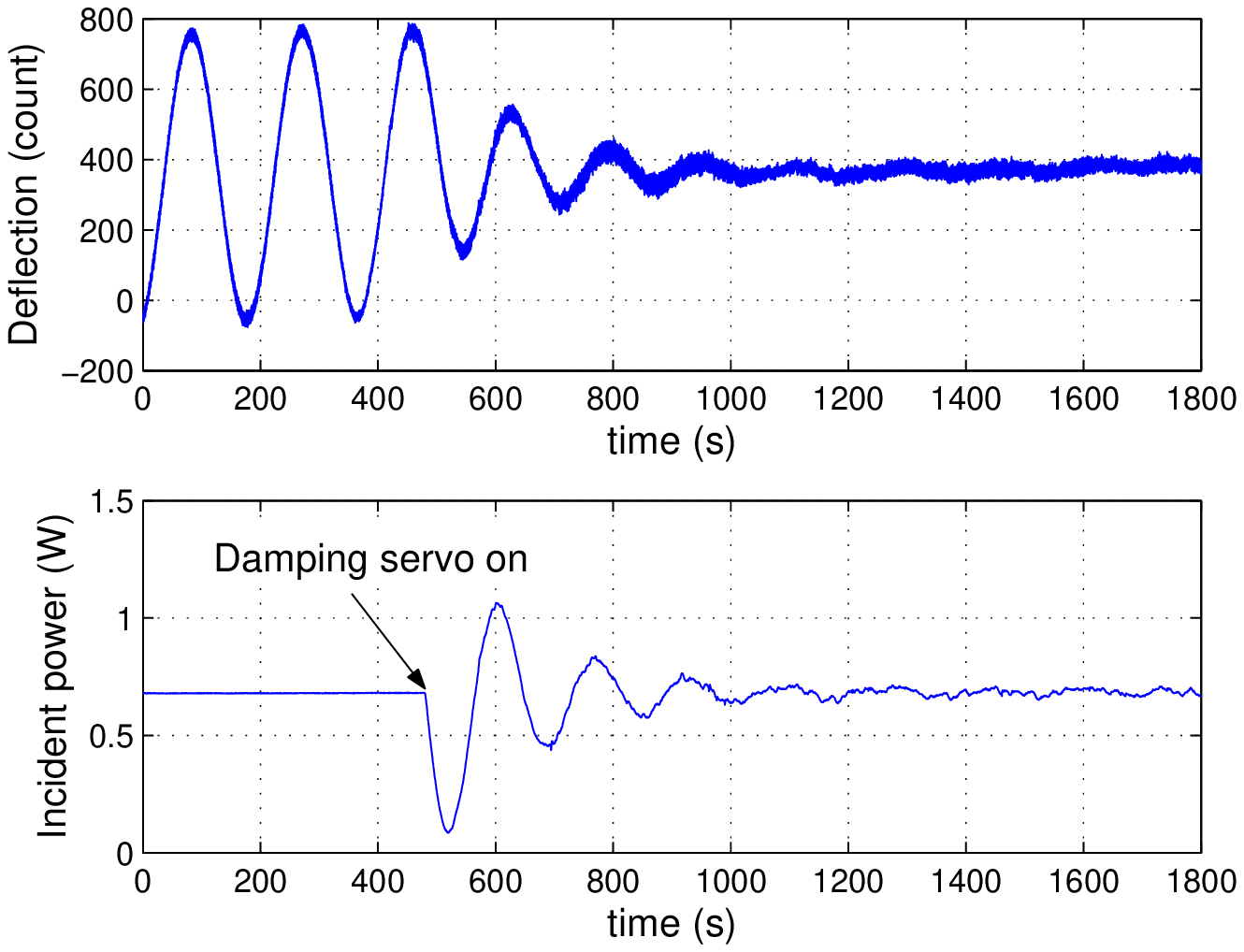}
\caption{Damping created by radiation pressure force.  
Upper:  deflection measured by the interferometer, 1 count corresponds to 10 nm.  Lower:  incident power on the mirror M4.  
After the damping servo switched on, the incident power was varied to generate torque needed to damp the pendulum's motion.}
\label{damping_t}
\end{center}
\end{figure}

Figure \ref{damping_t} is a time-domain plot showing that the pendulum is effectively damped by the radiation pressure.
\begin{figure}
\begin{center}
\includegraphics[angle=0,scale=0.63]{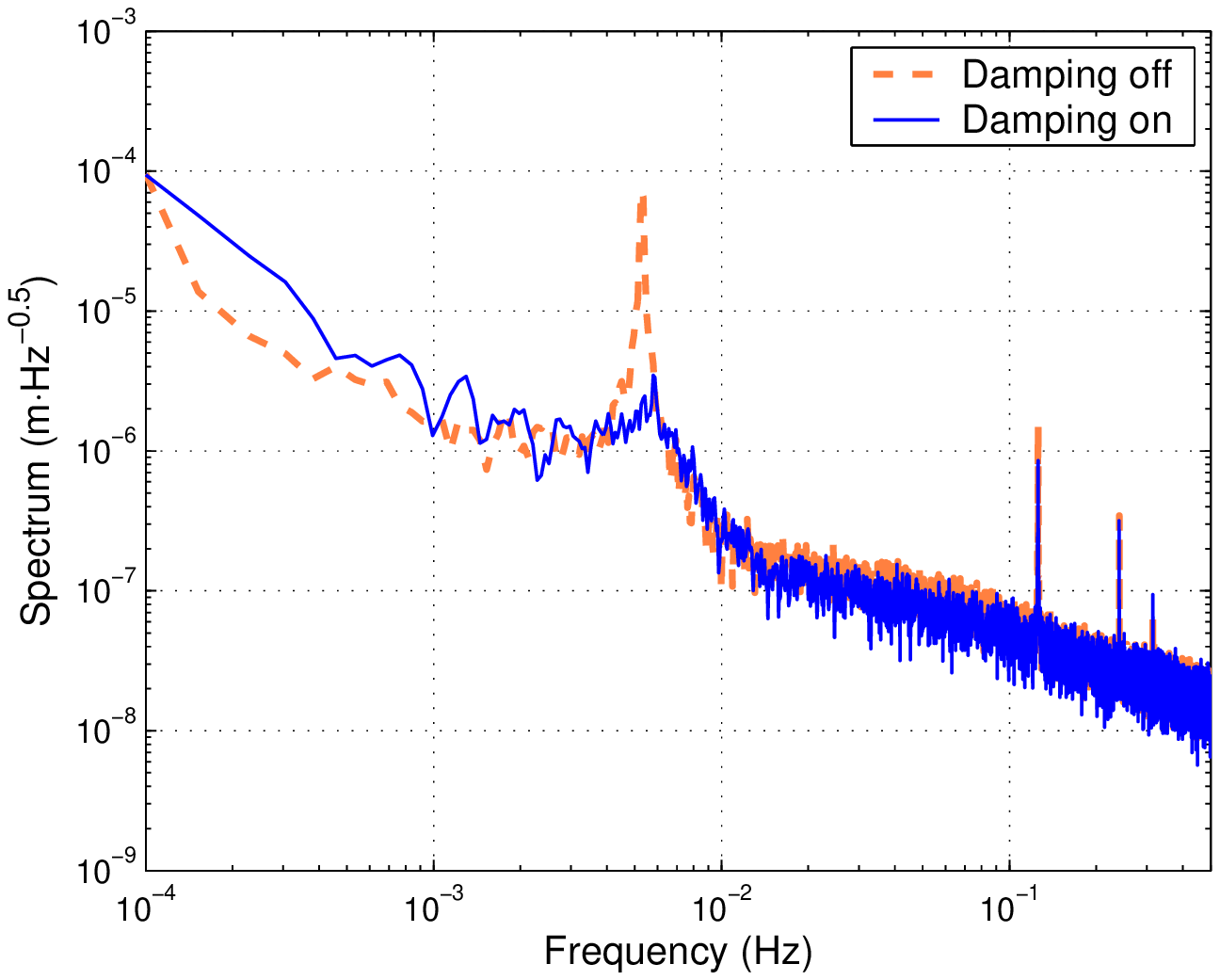}
\caption{The measured noise spectral density of the pendulum deflection.}
\label{damping_f}
\end{center}
\end{figure}
\begin{figure}
\begin{center}
\includegraphics[angle=0,scale=0.63]{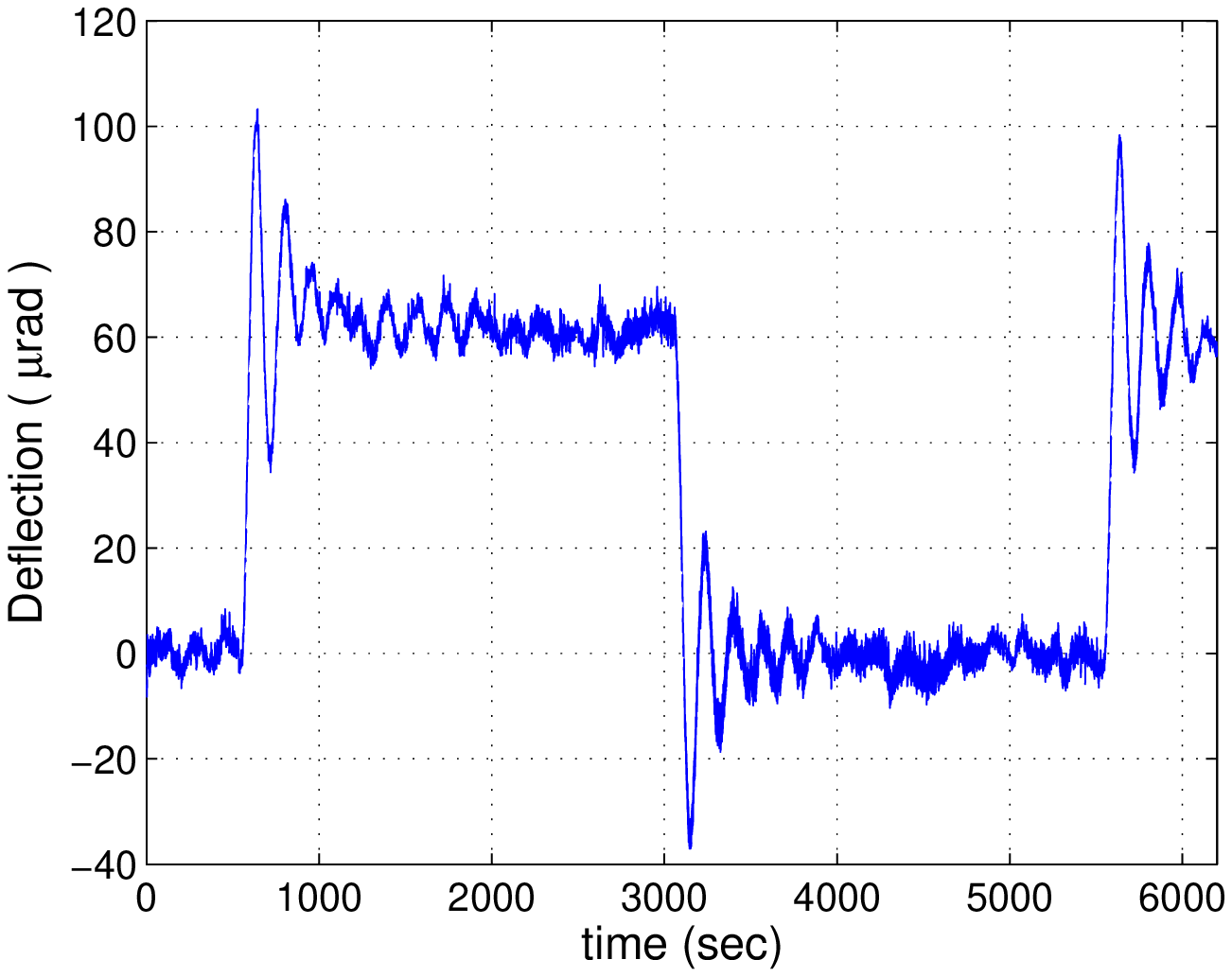}
\caption{Deflection excited by intensity modulated Nd:YAG laser.  The modulation is square wave with depth of 0.75 W.}
\label{rp_step}
\end{center}
\end{figure}
The corresponding Q factor is decreased from 272 to 3.4.  
This decrease can also be seen in the measured noise spectral density of pendulum deflection as shown in figure \ref{damping_f}.  

To quantitatively measure the force generated by the radiation pressure, 
the DC bias is periodically switched between 0.3 W and 1.05 W.  
According to (\ref{F_r}), this corresponds to a force variation of 5 nN.  
Figure \ref{rp_step} shows one segment of measured deflection due to the bias modulation.  
The mean deflection excited by the Nd:YAG laser is found to be $66\pm6$ $\mu\mbox{rad}$.  
Converting to force by 70 mm torque armlength and the spring constant of the pendulum, 
the force is $4.9\pm0.4$ nN which is in good agreement with the prediction.  
The light source for deflection measurement is a 633 nm laser with its maximum power close to 1mW, 
and therefore has negligible influence on pendulum deflection.    
The damping control is successfully demonstrated and will be used in the measurement of Newtonian constant of gravity in our future works.  \\

\section{Capacitive displacement sensing/actuating unit}

\begin{figure}
\begin{center}
\includegraphics[angle=0,scale=0.55]{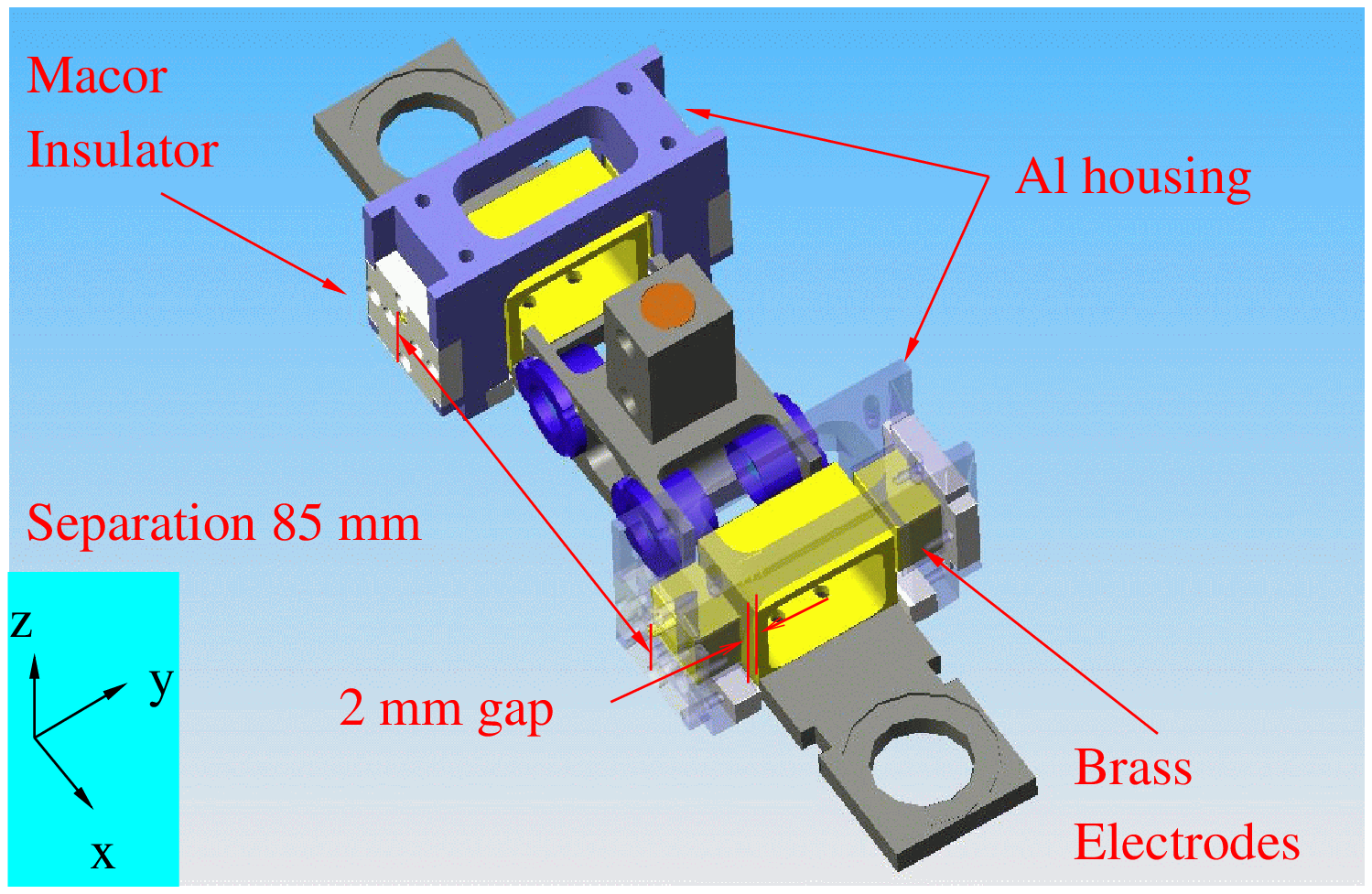}
\caption{CAD drawing of capacitive displacement sensing/actuating prototype.}
\label{pendulum_n_cap}
\end{center}
\end{figure}

\begin{figure}
\psfrag{DeltaC}[][][1.0][0]{$\Delta C$ (V)}
\begin{center}
\includegraphics[angle=0,scale=0.6]{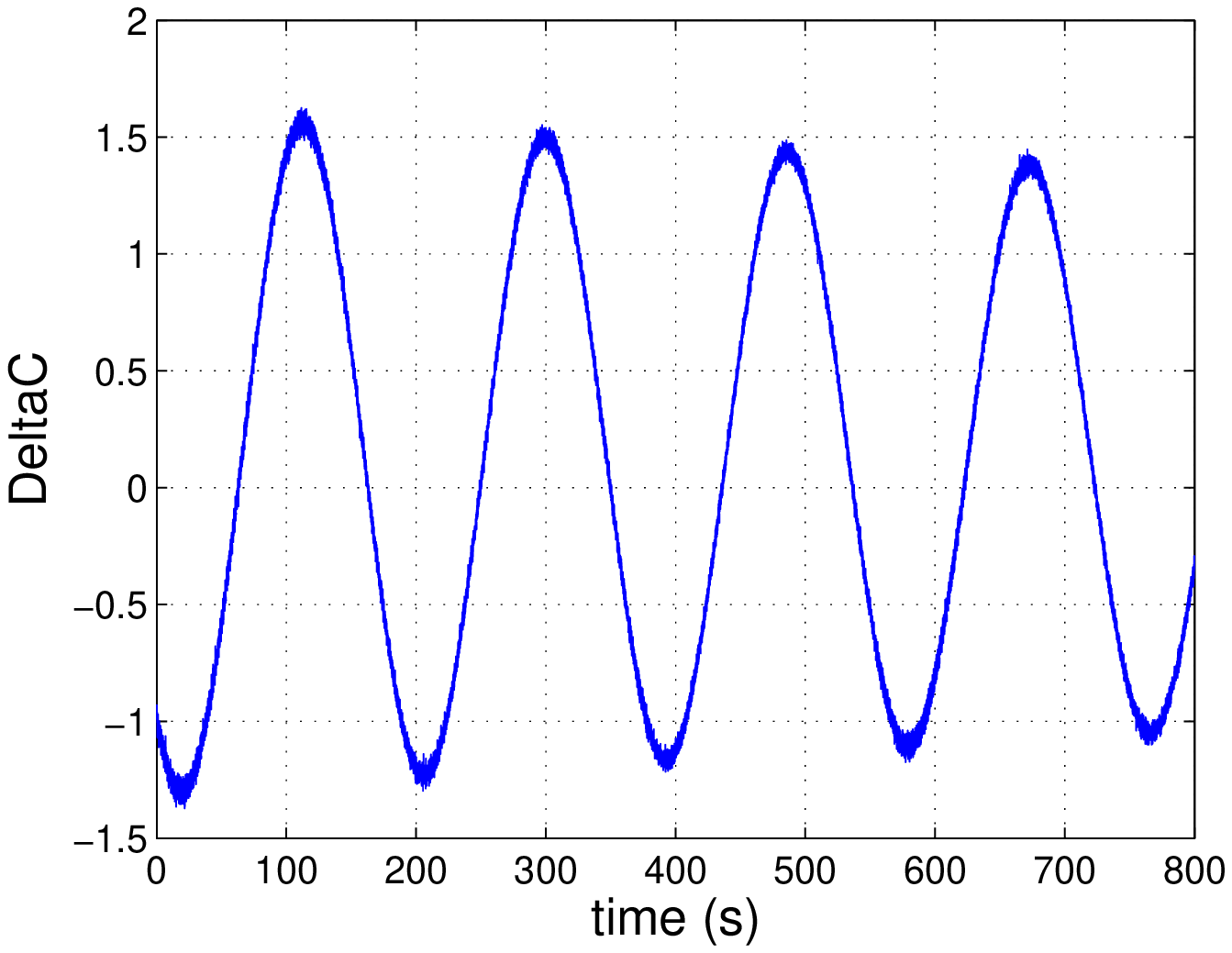}
\caption{Capacitance variation resulting from the pendulum motion.}
\label{cap_sensing_data}
\end{center}
\end{figure}

A new design of torsion pendulum equipped with a capacitive displacement sensing/actuating system is being tested now.  
Its key feature is to keep the torsion pendulum at its null position and compensate any forces acting on the pnedulum.  
Another advantage is that the null position is defined by the sensor itself, 
while with the optical interferometer, an offset has to be set and continuous interrogation is unavoidable.  
The prototype design is shown in figure \ref{pendulum_n_cap}.  
Here, we adopted the design concept of the capacitive accelerometer widely used in fundamental physics space experiments \cite{Josselin,Weber}.  
Each displacement sensing/actuating unit consists of three electrodes, 
where the middle one is mounted on the pendulum and the other two are fixed on the aluminum housing via Macor spacer.  
When the pendulum is at its equilibrium position, 
the gap distances of between two side electrodes and the middle electrode are the same, 
making capacitance values between electrodes equal.  
Hence the displacement of pendulum can be detected by differential capacitance sensing using a capacitive-inductive bridge circuit \cite{Josselin,Weber}.  
Figure \ref{cap_sensing_data} shows the preliminary displacement data of capacitance variation resulting from the pendulum motion.  
The oscillation frequency estimated by this data is 5.3 mHz, showing the correct functionality of the prototype design.  
Further work will be focused on the calibration and sensitivity improvement of the capacitive displacement sensor.  


\section{Future works}
Our next step is to calibrate the radiation force with the gravitation force.  
The torque from the radiation force will be balanced by the gravitational torque.  
This involves the precision measurement of gravitational constant G and the fine tuning of rotation table on which source masses are seated.  
The calibration also provide another way to trace the optical power to mechanical unit standards, 
the mass and length standards, through the gravitational constant G.  Hopefully, the first results will come out before the end of 2008.  

\section*{Acknowledgement}
This work is supported by Bureau of Standards, Metrology and Inspection (BSMI), Taiwan, R.O.C.






\begin{thebibliography}{10}
\bibitem{Newell} Newell D.B., Kramar J.A., Pratt J.R., Smith D.T. and Williams E.R., "The NIST microforce realization and measurement project", 2003, 
{\it IEEE Trans. Instrum. Meas.}, {\bf 52}, 508.  
\bibitem{Nesterov} Nesterov V., "Facility and methods for the measurement of micro and nano forces in the range below $10^{-5}$ N with a resolution of $10^{-12}$ N (development concept)", 2007, {\it Meas. Sci. Technol.}, {\bf 18}, 360-366.
\bibitem{Chen} Chen S-J, Pan S-S and Wu J-S, "Active damping control of a torsion pendulum by radiation pressure", 2008 Conference on Precision Electromagnetic Measurements Digest, Alan H. Cookson, Troy-Lee Winter, Editors, 132-133.
\bibitem{Quinn_1} Quinn T. J., Speake C. C., Richman S. J., Davis R. S. and Picard A., "A New Determination of G Using Two Methods", 2001, {\it Phys. Rev. Lett.}, {\bf 87}, 111101.  
\bibitem{Quinn_2} Quinn T. J., Davis R. S., Speake C. C. and Brown L. M., "The restoring torque and damping in wide Cu-Be torsion strips", 1997, {\it Phys. Lett.}, {\bf A228}, 36-42.
\bibitem{Wu} Wu J-S, Yeh H-C and Pan S-S, "Measurement of photon pressure", Proceedings of XVIII IMEKO WORLD CONGRESS, Metrology for a Sustainable Development, September, 17-22, 2006, Rio de Janeiro, Brazil.
\bibitem{Cohadon} Cohadon P. F., Heidmann A. and Pinard M., "Cooling of a Mirror by Radiation Pressure", 1999, {\it Phys. Rev. Lett.}, {\bf 83}, 3174.
\bibitem{Feat} Feat M., Zhao C., Ju L. and Blair D. G., "Demonstration of low power radiation pressure actuation for control of test masses", 2005, {\it Rev. Sci. Inst.}, {\bf 76}, 036107.
\bibitem{Metzger} Metzger C. H. and Karrai K., 2004, {\it Nature}, {\bf 432}, 1002.
\bibitem{Kleckner} Kleckner D. and Bouwmeester D., "Sub-kelvin optical cooling of a micromechanical resonator", 2006, {\it Nature}, {\bf 444}, 75.
\bibitem{Josselin} Josselin V., Touboul P. and Kielbasa R., "Capacitive detection scheme for space accelerometers applications", 1999, {\it Sensors and Actuators}, {\bf 78}, 92-98.
\bibitem{Weber} Weber J. W. {\it et al}, "Position sensors for flight testing of LISA drag-free control", 2003, Gravitational-Wave Detection, Mike Cruise, Peter Saulson, Editors, Proceedings of SPIE Vol.4856 31-42.
\end{thebibliography}
\end{document}